\documentstyle[12pt]{article}

\begin{document}

\newcommand{\bq}{{\bf q}}
\newcommand{\bmq}{{\bf |q|}}
\newcommand{\bk}{{\bf k}}
\newcommand{\bmk}{{\bf |k|}}
\newcommand{\bkq}{{\bf kq}}
\newcommand{\w}{\omega}
\newcommand{\W}{\Omega}

\begin{titlepage}
\begin{flushright}
{December 1998}
\end{flushright}
\vskip 0.5 cm

\begin{center}
            {\large{\bf A CASE OF WELL-DEFINED THERMAL DERIVATIVE EXPANSION TO
LOWEST ORDER}}
            \vskip 0.6 cm
 Marcelo Hott\footnote{E-mail address: m.hott1@physics.oxford.ac.uk}\renewcommand{\thefootnote}{\fnsymbol{footnote}}\footnote{On leave of absence from UNESP - Campus de Guaratinguet\'a - SP - Brazil.}
, Georgios Metikas\renewcommand{\thefootnote}{\arabic{footnote}}\addtocounter{footnote}{-1}\footnote{E-mail address: g.metikas1@physics.oxford.ac.uk} \\

\vskip 0.3 cm
{Department of Physics, Theoretical Physics,\\
University of Oxford, 1 Keble Road, Oxford OX1 3NP}
\end{center}
\vskip 2 cm

\abstract{We examine  a very simple model for which the leading contribution to the
one-loop effective potential at finite temperature is uniquely defined
despite the presence of the Landau terms. In addition we report on the
usual non-analyticity at finite temperature in order to compare our
perturbative results with exact ones obtained in the
literature. Finally, we point out the significance of our conclusions
in the context of symmetry restoration at finite temperature.}    
  
\end{titlepage}

\section{Introduction}

It is well-known that for most of the theories and models the
effective action displays a
non-analytic behaviour at finite temperature
\cite{WeldonMishaps,Das1} and this puts in jeopardy the construction of
an effective potential based on the derivative expansion technique
\cite{Zuk,Hott}. Historically this problem was first pointed out in
the BCS theory context by Abrahams and Tsuneto \cite{Tsuneto} and
later was also seen to appear in the gluon \cite{Kalashnikov} and in
the photon self-energy \cite{WeldonCovariant} for example.

The reason for this behavior is that temperature effects give rise to Landau terms and these are responsible for the development of a new branch cut in the complex plane of the external momenta with a branch point at the origin, besides the one already present at zero temperature. Then we have two branch cuts, the usual one defined by

\[ {q_{0}}^{2} - {\bmq }^{2} \geq  4m^{2} \]

\noindent and another which present only at finite temperature, namely

\[ {q_{0}}^{2} - {\bmq}^{2} \leq  0 \].

This second branch cut is not common to all graphs at finite
temperature. In fact it was shown that whenever the internal
propagators in a typical loop have different masses the branch point
is not in the origin, and the branch cut extends from $(m_{1} -
m_{2})^2$ to $-\infty $ \cite{vokos} where $m_1$ and $m_2$ are the masses of the particles in the internal loop.

In this letter we present another model where at least one of the
graphs displays analytic behaviour at zero external momenta. We show in
two different ways that the limit is well-defined and explain why this
is so. We also show how the non-analyticity can develop
in another model and compare our results with those obtained using non-perturbative techniques.  

 In section 2, we present the model  and examine carefully the
behaviour of the boson self-energy bubble diagram as the external
momenta go to zero. In section 3, we replace the parity
conserving interaction term of the theory with a similar but parity
violating one and examine the consequences.

 \section{A new  case} 

 We consider the following model
 \begin{equation}
 L[\bar{\psi}, \psi, \phi] = \bar{\psi} (i \not\! \partial - m) \psi -
 ig \bar{\psi} \gamma_5 \psi \phi + L_0[ \phi ] \label{L} 
 \end{equation}

 \noindent where $L_0[\phi]$ is the free Klein-Gordon Lagrangian. The boson is taken to be a pseudo-scalar quantity. 

We consider $\phi(x)$ to be an external field and
we want to obtain  the one loop contribution to the effective action
which is given by 
 \begin{equation} \Gamma_{eff} [\phi]= 
 -i \ln{\frac{ { \mathrm{Det}} [i {S}^{-1}[\phi]]}{ { \mathrm{Det}} [i S^{-1}]}}  \label{Det}
 \end{equation}
 \noindent where $i S^{-1}[\phi]$ and $i {S}^{-1} $ are matrices whose
elements in coordinate representation are  
 \begin{eqnarray*}
 \langle x|iS^{-1}|y \rangle &=& (i \not\! \partial_x -m) \delta(x-y) \\
 \langle x|i {S}^{-1}[\phi]|y \rangle &=&(i \not\! \partial_x -m -
ig \gamma_5 \phi (x)) \delta(x-y) 
 \end{eqnarray*}

 \noindent Since the external field depends on the coordinates, the
 resulting functional determinants are not straightforward to
 calculate. The matrices whose functional determinants we want to evaluate
  are not diagonal in momentum or in coordinate space. However,  
  we can write equation (2) as 
 \begin{equation}
\Gamma_{eff}[\phi ] =
 {-i \mathrm{Tr} }  \ln{ [ 1-ig \gamma_5 (-i S) {\phi}(x)]}.  
 \end{equation}
 \noindent Now we expand it in powers of the coupling constant and
show that the leading  contribution to the
 one-loop effective action is 

\[ \Gamma^{(2)} = \frac{ig^2}{2} \int \frac{d^4q}{(2 \pi)^4}
 \tilde{\phi}(-q) i \Pi (q) \tilde{\phi}(q) , \]

\noindent where
\begin{equation}
 i \Pi (q) = - \int \frac{d^4k}{(2 \pi)^4} {\mathrm{tr}}  \left[ \gamma_5 \frac{1}{\not\! k 
  + \not\! q - m} \gamma_5 \frac{1}{ \not\! k -m} \right] 
 \end{equation}

 \noindent  We note that $i \Pi(q) $ is just the self-energy
 bubble diagram for the boson which, after performing the trace, is given by  
 \begin{equation}
 i \Pi(q)= 4 \int \frac{d^4k}{ (2 \pi)^4 }
 \frac{ k^2 + k^{\mu}q_{\mu} - m^2}{[(k+q)^2 - m^2][ k^2 - m^2]} . \label{new}
 \end{equation}
\noindent This is one typical diagram that usually has a
non-analytical behavior in the limit of vanishing external momenta,
but we are going to show that this is not the case here. We  keep
this intermediate expression because it will help us to show in the
next section how the
non-analyticity can develop in the scalar-coupling model.

It is worth mentioning that the leading contribution to the one-loop
effective action can also be written as

\[ \Gamma^{(2)} = \frac{ig^2}{2} \int {d^4 x}
 \; \phi (x) i \Pi (q) \phi (x) , \]

\noindent where $q_\mu$ is to be understood as a derivative operator
acting on the field to the right. This result can be obtained by
inserting complete sets of eigenvectors of momentum and position
operators. The external field is an operator acting on coordinate
states which at finite temperature are thermal states. Then the
eigenfunction $\phi(x)$ has to be interpreted as being thermalized
itself. This seems to us to be the reason why the approach given in
\cite{EvansDerivative,EvansUnique} is not in agreement with what is
usually expected for the effective action at finite temperature.
 
Applying the usual finite temperature techniques to (\ref{new}), we find the following
 expression for the thermal bubble diagram.
 \begin{eqnarray}
 &&  \Pi (q_{0}, {\bf q}) = - \int \frac{d^3 \bk }{ (2 \pi )^3 } \
 \frac{1}{ 2 \w \W } \left\{ 4 \w \tanh{ \frac{  \beta \W }{2}} +
\right. \nonumber \\
 &&+ \frac{1}{ \W + \w - q_0 } \left[ [ \w q_0 + \bkq ] \tanh{ \frac{
 \beta \w }{2} } + [ q_0^2 - \W q_0 + \bkq ] \tanh{ \frac{\beta \W }{2}} \right]   \nonumber \\ 
 &&+ \frac{1}{ \W + \w + q_0 }
  \left[ [ - \w q_0 + \bkq ] \tanh{ \frac{ \beta \w }{2}} + [q_0^2 + \W q_0
 + \bkq ] \tanh{ \frac{ \beta \W }{2}} \right] \nonumber \\
 &&+ \frac{1}{ \W - \w + q_0 }
 \left[ [ \w q_0 + \bkq ] \tanh{ \frac{ \beta \w }{2}} - [ q_0^2 + \W q_0
 + \bkq ] \tanh{ \frac{ \beta \W }{2}} \right] \nonumber \\
 && \left. + \frac{1}{ \W - \w - q_0 }
 \left[ [- \w q_0 + \bkq ] \tanh{ \frac{ \beta \w }{2}} - [ q_0^2 - \W q_0
 + \bkq ] \tanh{ \frac{ \beta \W }{2}} \right] \right\}, \nonumber \\  
\label{thermal bubble}
 \end{eqnarray}

 \noindent where \[ \w = \sqrt{ \bk ^2 + m^2} \hspace{3cm} \W = \sqrt{
( \bk + \bq )^2 + m^2 } \; . \]

We can have a first indication that the zero-momentum limit of
expression (\ref{thermal bubble}) does not display the usual
non-uniqueness problem by examining the two successive limits. Namely,

\begin{eqnarray*}
\Pi (0, {\bf q}) & = &  - \int \frac{d^3 \bk }{ (2 \pi )^3 } \; \frac{1}{
\w \W } 
\left\{ 2 \w \tanh{\frac{  \beta \W }{2} }  
+  \frac{ \bkq }{ \W + \w } \left[ \tanh{\frac{ \beta \w }{2} } +
\tanh{\frac{ \beta \W }{2} } \right] \right.  \nonumber \\ 
& + &  \left.  \frac{\bkq}{\W - \w }
\left[ \tanh{ \frac{ \beta \w }{2}} - \tanh{\frac{ \beta \W }{2}} 
\right] \right\} , 
\end{eqnarray*}

\noindent which can be checked to give
 
\begin{equation}
 \lim_{ {\bmq} \rightarrow 0} \Pi (0, {\bf q}) = -\frac{1}{{\pi}^2}
\int_{|m|}^{ \infty }
 d \w \sqrt{{\w}^{2}-m^{2}}\tanh{ \frac{ \beta \w }{2}} \; .
\end{equation}

\noindent Similarly we find

\begin{eqnarray*}
  \Pi (q_{0}, 0) & = & - \int \frac{d^3 \bk }{ (2 \pi )^3 } \
 \frac{1}{ 2 \w \W } \left\{ 4 \w \tanh{ \frac{  \beta \W }{2}} +
\right. \nonumber \\
 & + & \left. {q_{0}}^{2} \left[ \frac{1}{ 2 \w - q_0 } + \frac{1}{ 2 \w +
q_0} \right] \tanh{ \frac{ \beta \w }{2}}  \right\}.  \nonumber \\
& \stackrel{q_{0} \rightarrow 0}{ \longrightarrow } &    -\frac{1}{{\pi}^2}
\int_{|m|}^{ \infty } d \w \sqrt{{\w}^{2}-m^{2}} \tanh{ \frac{\beta
\w}{2}} \; .
\end{eqnarray*}

\noindent We conclude that the limits coincide. Moreover, the only term
that contributes to the unique result is the first one in the
integrand of equation (\ref{thermal bubble}) and those proportional to Landau terms -
the last two terms inside the integrand - vanish in this limit. 

A more general way of seeing that the limits are the same is to
perform the
parameterization $ q_0 = a \bmq $, where $a$ can be
 any real number, and find the limit of $ \Pi (a \bmq , \bmq ) $ as $
\bmq \rightarrow 0 $. If the limit is independent of $a$, we have a
strong indication that the function
is analytic at the origin, {\it i.e.} it does not depend on the way one
approaches the origin \cite{WeldonMishaps}. Before doing so we recast
equation (\ref{thermal bubble}) in a more convenient form by means of
the transformation $ \bk \rightarrow  - (\bk + \bq) $ wherever the integrand contains $\tanh{\frac{\beta \W}{2}} $. Then
we find  

\begin{eqnarray}
 \Pi & = & - \int \frac{d^3 \bk }{ (2 \pi )^3 } \left\{ \frac{2}{ \w }
  \tanh{ \frac{\beta \w }{2} } + ( q_0^2 - \bq ^2 ) \tanh{ \frac{ \beta \w
 }{2} } \right. \nonumber \\
 & \times & \left.\frac{1}{2 \w \W} \left[ \frac{1}{ q_0 + \W + \w } - \frac{1}{
 q_0 - \W - \w } + \frac{1}{ q_0 + \W - \w } - \frac{1}{ q_0 - \W + \w
 } \right] \right\}
 \end{eqnarray}
 \noindent One can note that at $T=0$ the Landau terms cancel each other, as expected.
 We change variables from $ \cos{\theta} $ to $ \W $ and perform the
 integration over $ \W $. The result is 

\begin{eqnarray}
\Pi ( a\bmq , \bmq ) & = &  -\frac{1}{{\pi}^2}
\int_{|m|}^{ \infty } d \w \sqrt{{\w}^{2}-m^{2}} \tanh{ \frac{ \beta\w}{2} } -  \nonumber \\
& - &  \frac{ (a^2 - 1)\bmq ^2 }{2 \bmq } \int_{|m|}^{ \infty } \frac{ d \w }{ (2
 \pi)^2 } \tanh{ \frac{\beta \w}{2} } [ L1 + L2 + L3+ L4 ] 
 \end{eqnarray}
 
\noindent where 
 \begin{eqnarray*}
 L1( \bmq ) = \ln{ \frac{ \W _{+} + \w + a\bmq }{ \W_{-} + \w +  a\bmq}} &&
 L2( \bmq ) = \ln{ \frac{ \W _{+} + \w - a\bmq  }{ \W_{-} + \w -  a\bmq}} \\
 L3( \bmq ) = \ln{ \frac{ \W _{+} - \w +  a\bmq }{ \W_{-} - \w +  a\bmq}} &&
 L4( \bmq ) = \ln{ \frac{ \W _{+} - \w -  a\bmq }{ \W_{-} - \w -  a\bmq}} 
 \end{eqnarray*}
 
\noindent with \[ \W _{+}= \sqrt{ (\bmk + \bmq )^2 + m^2 }
 \hspace{2cm}  \W _{-}= \sqrt{ (\bmk - \bmq )^2 + m^2 }. \]

\noindent  The limits of two of the regular terms $L1$ and $L2$ are independent
of $a$ as they should be. We can see that

 \[ \lim_{ \bmq \rightarrow 0} (a^2 -1)\bmq L1 = 0 \hspace{2cm} \lim_{ \bmq
\rightarrow 0} (a^2 -1)\bmq L2 = 0 . \]

\noindent  What is quite unexpected is that, for this particular
model, the contributions coming from the Landau terms, $L3$ and $L4$,
vanish independently of $a$, that is

 \[ \lim_{
 \bmq \rightarrow 0 } (a^2 -1)\bmq L3 = 0   
  \hspace{2cm} \lim_{ \bmq \rightarrow 0} (a^2 -1)\bmq L4 = 0 .\]

\noindent In other
 words, although the Landau terms are not well-behaved at the origin of momentum space a unique effective potential can be
 defined thanks to the kinetic term in the numerator of equation (8),
namely  $q_0^2 - {\bf q}^2$. This is an interesting result, but this
kinetic term does not always appear in bubble diagrams as we are going to
see in the next section.  
  In the present case the one-loop, $g^2$ order contribution to the effective potential is 
 \begin{eqnarray}
 V_{eff}^{(2)} &=& - \frac{i g^2}{2} i \Pi (0,0)  \phi^{2} \nonumber \\
 \Pi (0,0) &=& -\frac{1}{ \pi ^2} \int_{|m|}^{ \infty } d \w \sqrt{ \w ^2
 - m^2 } \tanh{ \frac{ \beta \w }{2} } \; .  
 \end{eqnarray}
 
The next order in the derivative expansion is non-analytic since the
derivatives of the Landau terms become dominant and the derivative expansion breaks down.
 
 \section{ A Usual Case }
 If we replace in equation (\ref{L}) the interaction term with one
which does not
 contain the $ \gamma_5 $ matrix and repeat the same procedure we find
 
\begin{equation}
 i \Pi^{'}(q)=  4 \int \frac{d^4k}{ (2 \pi)^4 }
 \frac{k^2 + k^{\mu}q_{\mu} + m^2}{[(k+q)^2 - m^2][ k^2 - m^2]} \label{usual}
 \end{equation}
 
\noindent which can be written as

\[ i \Pi^{'}(q) =  i \Pi(q) + i \Pi^{''}(q), \]

\noindent where

\begin{eqnarray*}
 i \Pi^{''}(q) = 4 \int \frac{d^4k}{(2 \pi)^4} \frac{2 m^2}{[(k+q)^2 -
m^2][ k^2 - m^2]} \; .
\end{eqnarray*}	     

As we saw in the previous section $ \Pi(a\bmq, \bq)$ does not depend
on $a$, when $\bq \rightarrow 0$. On the other hand $\Pi^{''}(q) $ does. In
fact, we have

\begin{eqnarray}
&& \lim_{\bmq \rightarrow 0} \Pi^{''}(a \bmq ,\bmq ) =  \frac{m^2}{\pi^2}
\int_{|m|}^{ \infty } d \w \left\{ \frac{ \sqrt{ \w^2 - m^2 } }{\w^2} \tanh{
 \frac{ \beta \w}{2} } - \right. \nonumber \\
&& -  \left. \frac{\beta}{2 \w}    \cosh^{-2}{ \frac{\beta \w}{2} }
\left[ \sqrt{\w^2 - m^2} - \frac{\w a}{2} \ln{ \frac{| \w a + \sqrt{
\w^2- m^2 }| }{|\w a - \sqrt{ \w^2- m^2 }| } } \right] \right\} .  
\end{eqnarray}	     

\noindent We see that dropping  $\gamma_5$ from the interaction has
 made a great difference which is reflected on the relative signs
of the terms in the numerator (Compare equation (\ref{new}) to
equation (\ref{usual}).).

It is important to compare our results with other in the
literature and, in particular, with Dolan and Jackiw \cite{dolan} . For the one-loop effective potential at order $g^2$ we
have

\begin{eqnarray}
&& V^{(2)}_{eff} =  - \frac{g^2}{2 \pi^2} \int_{|m|}^{\infty} d\w \left\{
\sqrt{\w^2 - m^2} \left( 1 - \frac{m^2}{\w^2} \right)
\tanh{ \frac{ \beta \w }{2} } \; + \right. \nonumber \\
&& \left. \frac{\beta }{2 \w } \cosh^{-2}{\frac{\beta \w }{2} } \left[
\sqrt{\w^2 - m^2} - \frac{\w a}{2} \ln{ \frac{| \w a + \sqrt{ \w^2-
m^2}| }{| \w a - \sqrt{\w^2- m^2} | } } \right] \right\} \;
\phi^{2}(x) .  \label{our}
\end{eqnarray}

\noindent This gives the contribution for the thermal mass of the $\phi(x)$
field. Considering the external field to be constant Dolan and Jackiw obtained the following exact expression for the
one-loop effective potential

\begin{equation}
V_{eff}= - \frac{2}{\pi^2} \int_{|m|}^{\infty} d\w \sqrt{\w^2
-m^2} \left[\frac{E}{2} + \frac{1}{\beta} \ln{(1- e^{\beta E})}
\right], \label{exactveff}
\end{equation}

\noindent{where}

\[ E= \left[ \sqrt{\w^2 - m^2} + (m + g \phi)^2 \right]^{1/2}. \] 

\noindent We are interested in the contribution at the second order in the
coupling constant which is

\begin{eqnarray}
V^{(2)}_{eff} & = & - \frac{g^2}{2 \pi^2} \int_{|m|}^{\infty} d\w
\sqrt{\w^2 - m^2} \left\{ \left( 1 - \frac{m^2}{\w^2} \right)
\tanh{\frac{\beta \w}{2} }+ \right. \nonumber \\
& + & \left. \frac{ m^2 \beta}{2 \w}  \cosh^{-2}{ \frac{\beta \w }{2}
} \right\} \; \phi^{2} . \label{dolan}
\end{eqnarray}

\noindent If we set $a=0$ in equation (\ref{our}) it reduces to expression
(\ref{dolan}), which means that the result derived by Dolan and Jackiw
is valid only in one of the infinite number of ways of approaching the
origin, namely in the case where we take $q_0 \rightarrow 0 $ first
and then $ \bq \rightarrow 0$. One can also
reproduce equation (\ref{dolan}) by setting $(q_0, \bq ) = (0, 0) $ in formula (\ref{usual}) and then performing the Matsubara
sum. This is also equivalent to assuming from the beginning that the
derivative expansion is well defined at finite temperature. However,
the correct thing to do is to perform the sum first and then see how
$\Pi (q) $ behaves in the limit $(q_0, \bq ) \rightarrow  (0, 0) $. We
therefore conclude that the non-perturbative method employed in
\cite{dolan} is not
generally equivalent to the perturbative calculation because it fails
to take into account the non-analyticity which appears at the origin
of the space of external momenta.  

\section{Conclusions}

We have shown that in a  model, where  fermions couples to a
pseudo-scalar field, the thermal mass for the pseudo-scalar field can be
found uniquely at finite temperature. We have also shown that this is
not true when the fermion couples to a scalar field, the reason for
that being  the non-analytic behaviour which appears in the thermal
bubble diagram. The models we dealt with can be
considered together to study chiral symmetry restoration at finite
temperature for example in the linear $\sigma$ model \cite{Gell} in its broken
chiral symmetry phase and in the Nambu-Jona-Lasinio model
\cite{NJL} expressed in terms of auxiliary fields.  

Finally, we point out that, whenever finite temperature chiral symmetry
restoration is discussed by employing non-perturbative results for the effective
potential, they may not match those based on perturbation
theory. Therefore the question of symmetry restoration at finite
temperature should be reanalyzed keeping in mind the non-analyticity
of some graphs. Work on this and other related issues is in progress.

\newpage
\section*{Acknowledgments}
M. Hott is supported by Funda\c c\~ao de Amparo a Pesquisa do Estado de
S\~ao Paulo (FAPESP-Brazil). G. Metikas is grateful to PPARC (UK) for
financial support. The authors wish to thank Prof. I.J.R Aitchison for
valuable discussions.   



\end{document}